\documentclass[12pt,preprint]{aastex}

\newcommand{\gsim}{~\rlap{$>$}{\lower 1.0ex\hbox{$\sim$}}}
\newcommand{\lsim}{~\rlap{$<$}{\lower 1.0ex\hbox{$\sim$}}}

\begin{document}

\title{The luminosity functions and stellar masses of galactic disks
and spheroids}  

\author{A.~J.~Benson\altaffilmark{1}
C.~S.~Frenk\altaffilmark{2} \& R.~M.~Sharples\altaffilmark{2}}

\affil{1. California Institute of Technology, MC 105-24, Pasadena, CA 91125,
U.S.A.\\ (e-mail: abenson@astro.caltech.edu)}

\affil{2. Physics Department,
University of Durham, Durham, DH1 3LE, England}

\begin{abstract}

We present a method to obtain quantitative measures of galaxy morphology
and apply it to a spectroscopic sample of field galaxies in order to
determine the luminosity and stellar mass functions of galactic disks and
spheroids. For our sample of approximately 600 galaxies we estimate, for
each galaxy, the bulge-to-disk luminosity ratio in the I-band using a
two-dimensional image fitting procedure. Monte Carlo simulations indicate
that reliable determinations are only possible for galaxies approximately
two magnitudes brighter than the photometric completeness limit, leaving a
sample of 90 galaxies with well determined bulge-to-total light
ratios. Using our measurements of individual disk and bulge luminosities
for these 90 galaxies, we construct the luminosity functions of disks and
spheroids and, using a stellar population synthesis model, we estimate the
stellar mass functions of each of these components. The disk and spheroid
luminosity functions are remarkably similar, although our rather small
sample size precludes a detailed analysis. We do, however, find evidence in
the bi-variate luminosity function that spheroid-dominated galaxies occur
only among the brightest spheroids, while disk-dominated galaxies span a
much wider range of disk luminosities. Remarkably, the total stellar mass
residing in disks and spheroids is approximately the same. For our sample
(which includes galaxies brighter than $M_* +2$, where $M_*$ is the
magnitude corresponding to the characteristic luminosity), we find the
ratio of stellar masses in disks and spheroids to be $1.3 \pm 0.2$. This
agrees with the earlier estimates of Schechter \& Dressler, but differs
significantly from that of Fukugita, Hogan \& Peebles. Ongoing large
photometric and redshift surveys will lead to a large increase in the
number of galaxies to which our techniques can be applied and thus to an
improvement in the current estimates.

\end{abstract}

\section{Introduction}

It has long been known that galaxies in the nearby Universe display a
range of morphological characteristics that distinguish between
disk-dominated (e.g. spiral) and spheroid-dominated (e.g. elliptical)
galaxies, with many fine subdivisions within each class
(e.g. \citealt{hubble26}; \citealt{devac59}). The origins of these
different types of galaxy and the evolutionary connections, if any,
between them are still unclear, although there is a wealth of
observational data and several proposed theories.

Traditionally, morphology has been assigned by human classifiers
directly from galaxy images, a process that is accurate only to within
about two T-types \citep{naim95}. More recently, computer-based
algorithms have been developed which show a reasonable correlation
with ``eyeball'' estimates, at least for bright galaxies
(e.g. \citealt{abraham96}). Unfortunately, all these classifications
tend to give considerable weight to detailed morphological features,
such as spiral arms or asymmetries in the image, and are difficult to
compare with current theoretical predictions which focus on simpler
quantities such as the total stellar mass or luminosity in the disk
and spheroidal components. Here, we consider a morphological
quantifier that is more easily related to theoretical models. 

It is widely believed that disks form by slow accretion of gas which
acquired angular momentum through tidal torques
\citep{hoyle49,peebles69}, although whether this picture works in
detail remains an open question \citep{nfw95,navsteindummy,bosch01}.
Spheroids, on the other hand, are thought to form either by a
``monolithic collapse'' \citep{els,jimenez99}, or as a result of
mergers of pre-existing galaxies (\citealt{toomre77};
\citealt{barnes92} and references therein). Detailed theoretical
predictions for the statistical morphological properties of galaxies
and their evolution have been calculated for the hierarchical merging
formation mechanism appropriate to cold dark matter cosmologies
\citep{kauff93,kauff9596,baugh96ab,somerville01}. Among other things,
these models give the relative luminosities and stellar masses of the
spheroids and disks of galaxies.

In this work, we measure the I-band bulge-to-total light ratio, B/T,
for a large sample of galaxies with spectroscopic redshifts by fitting
two-dimensional models to the observed galaxy images. We use this
information to estimate the spheroid and disk luminosity functions, as
well as the total stellar mass that resides in disks and spheroids. An
earlier attempt to estimate the relative contributions to the
luminosity from spheroids and disks was carried out by
\citet{schecht87}, based on ``eyeball'' estimates of the
bulge-to-total ratio.

The remainder of this paper is arranged as follows. In \S\ref{sec:data} we
describe our basic dataset. In \S\ref{sec:method} we introduce the method
used to fit model images to the data and thereby extract B/T (along with
other interesting parameters) and describe how we estimate errors. In
\S\ref{sec:results} we examine the accuracy of our technique and
determine how well the B/T ratio can be measured as a function of the
apparent magnitude of a galaxy. We then compute luminosity functions
and total stellar masses for spheroids and disks. Finally, in
\S\ref{sec:disc} we present our conclusions.

\section{Data}
\label{sec:data}

The galaxy sample used in this work is that of \cite{gardner97}. The
reader is referred to that work for a full description of the
data. Here we summarize the most important features of the dataset.

Imaging of two fields of total area $10$ deg$^2$ was carried out in the B,
V and I bands using the T2KA camera on the Kitt Peak National
Observatory (KPNO) 0.9-m telescope, resulting in images with
$0.68$-arcsec per pixel. Exposures of 300s reached $5\sigma$ detection
depths of B=21.1, V=20.9 and I=19.6 in 10-arcsecond circular
apertures. Imaging was also carried out in the K-band using the IRIM
camera on the KPNO 1.3-m telescope resulting in 1.96-arcsecond per
pixel images, and a $5\sigma$ detection depth of K=15.6 in a
10-arcsecond circular aperture. The positions of the fields were
chosen randomly (the field centers are RA 14$^{\rm h}$15$^{\rm m}$,
Dec. $+00^\circ$ and RA 18$^{\rm h}$0$^{\rm m}$,
Dec. $+66^\circ$). The I-band images, which we will use in this work,
were bias-subtracted, flattened using twilight flats and with median
sky flats. Objects were identified with the {\sc sextractor} program
(\citealt{sex}), using a $3\sigma$ threshold. The seeing in the optical
images varied in the range $1.3 < {\rm FWHM} < 2.0$ arcsec. One field
contained a nearby rich galaxy cluster.

Spectroscopic follow-up was obtained for a K-selected sample in sub-regions
of total area $4.4$ deg$^2$, using the Autofib-2 fiber positioner and
WYFFOS spectrograph on the 4.2m William Herschel Telescope on La
Palma. Spectra were obtained for 567 galaxies with K$<15$, which allowed
redshifts to be measured for 510 galaxies (a redshift completeness of
90\%). Although the spectroscopic sample is K-selected, this does not
introduce any incompleteness in the I$<16$ sample used extensively in this
work (i.e. there are no galaxies with I$-$K$<1$ in the sample). We also
briefly consider an I$<18$ sample, for which the spectroscopic completeness
falls to around 50\% because of the K-band selection (which also introduces
a bias in this fainter sample against objects that are blue in I$-$K). For
the I$<16$ and I$<18$ samples the median redshift is $z=0.08$ and $0.14$
respectively.

\section{Method}
\label{sec:method}

\cite{wad99} have proposed a two-dimensional galaxy decomposition
technique that can efficiently recover B/T ratios (and other
parameters) of model galaxy images with high accuracy (see also
\citealt{byun95,dejong96}). They present a detailed study of the
effects of uncertainties in the point spread function (PSF), the
presence of nearby stars and the stability of the B/T estimates as a
function of signal-to-noise. Our approach is similar to theirs, but we
apply the technique to a large photometric sample of real
galaxies\footnote{\cite{wad99} applied their technique to three
galaxies for which previous estimates of B/T were available and found
reasonable agreement.}. We consider the ``real-world'' problems of
automated masking of nearby galaxies and stars and make a thorough
assessment of the errors in the measured parameters. We also present,
in an Appendix, estimators for the luminosity functions of disks and
spheroids.

For this analysis, we have used the sample of 636 ${\rm I}<18$
galaxies of \cite{gardner97} imaged in B, V, I and K, as described in
\S\ref{sec:method}. Postage stamp images of $33 \times 33$ pixels
($0.68^{''}$/pix) around each galaxy were extracted from the I-band
data. This was the best observed band by \cite{gardner97} and is
particularly well-suited for our purposes because it minimizes the
effects of young blue stars. This image size is large enough to
include the entire region of the galaxy for which reasonable
signal-to-noise is achieved (and in most cases extends well beyond
it.) To determine the bulge-to-total ratio, B/T, we fit the
two-dimensional surface brightness profile of each galaxy using a
combination of an exponential disk,
\begin{equation}
\Sigma_{\rm d}(\theta)=\Sigma_{\rm d,0} \exp (-\theta/\theta_{\rm d}),
\end{equation} 
and an $r^{1/4}$-law spheroid,
\begin{equation}
\Sigma_{\rm s}(\theta)=\Sigma_{\rm s,e} \exp
(-7.67[(\theta/\theta_{\rm e})^{1/4}-1]), 
\end{equation}
where $\theta$ is the angular distance from the galaxy center. We will
also consider a more general $r^{1/n}$ spheroid profile, as
\cite{wad99} did. The disk is allowed to be inclined, and to have
arbitrary position angle. The spheroid is allowed an ellipticity
(defined as the ratio of semi-major to semi-minor axes) in the range 1
to 6, and can also have an arbitrary position angle. To mimic seeing,
we construct mock images using these profiles which we then smooth
with a Gaussian filter (integrated over each pixel to account for the
variation of the PSF across the pixel), 
\begin{equation}
p(\theta)=\exp[-(\theta/\sigma)^2/2]/2\pi\sigma^2. 
\label{eq:see}
\end{equation}
The width of the Gaussian is treated as a free
parameter, to account for variations in seeing between the
images. (We will examine in \S\ref{sec:acc} the effect of using a
more realistic PSF.) The postage stamp images were centered on the
galaxy of interest, but we allow the position of the image center to
vary since in many cases the resulting sub-pixel variations lead to
lower values of $\chi^2$.  We also allow a small contribution from a
faint, constant surface brightness background in order to
take into account small inaccuracies in sky subtraction. The best
fitting parameters for each galaxy were then obtained by minimizing
$\chi ^2$ using Powell's algorithm \citep{brent73}. There are a total
of 12 fit parameters (13 if we include $n$ when fitting $r^{1/n}$
spheroid profiles) summarized in Table~\ref{tb:fitpars}. In
\S\ref{sec:acc} we will consider how accurately this procedure 
recovers the B/T ratio of the galaxies.

\begin{table}
\caption{The parameters used to construct mock galaxy images in the
fitting procedure. Each parameter is described in the text.}
\label{tb:fitpars}
\begin{center}
\begin{tabular}{cp{7cm}}
$\Sigma_{\rm d,0}$ & Disk central surface brightness \\
$\Sigma_{\rm s,e}$ & Spheroid surface brightness at the effective radius \\
$\theta_{\rm d}$ & Disk angular scalelength \\
$\theta_{\rm e}$ & Spheroid effective radius \\
$P_{\rm d}$ & Disk position angle \\
$P_{\rm s}$ & Spheroid position angle \\
$i$ & Disk inclination angle \\
$e$ & Spheroid ellipticity \\
$(x,y)$ & Center of image \\
$B$ & Excess background surface brightness \\
$\sigma$ & Seeing (c.f. eqn.~\protect\ref{eq:see}) \\
$n$ & Spheroid profile index (fixed at $n=4$ unless otherwise stated)
\end{tabular}
\end{center}
\end{table}

A significant fraction of the postage stamp images was contaminated by
a secondary galaxy (and occasionally by more than one). We use a
simple algorithm to identify such contaminants and mask them from the
image. The aim is to remove objects which are physically distinct from
the galaxy of interest without masking any pixels of the galaxy
itself. We first rank the pixels in the image by surface brightness
and then proceed to find groups of bright pixels. The brightest pixel
is assigned to the first group. Successive pixels are assigned to a
pre-existing group if they touch it (i.e. if they are adjacent either
horizontally, vertically or diagonally), or else are assigned to a new
group. In the case where a pixel touches more than one group, the two
groups are merged. This process is continued until pixels of a fixed
signal-to-noise are reached (specifically, we consider only pixels
more than $3\sigma$ above the sky background). If more than one group
exists at this point, the group at the center is deemed to be the
galaxy of interest and the pixels of all secondary groups are marked
as being contaminated and are not included in the $\chi ^2$ sum. This
simple algorithm works well in the majority of cases, but it fails in
a few (28 out of 636 galaxies), either by not removing a contaminating
galaxy, or by removing a significant fraction of the primary
galaxy. Rather than attempting to use a more complex algorithm in
these cases, we resorted to cleaning the image by hand (i.e. we view
the image and manually mark the contaminated pixels).

\clearpage

\begin{figure*}
\epsscale{.50}
\plotone{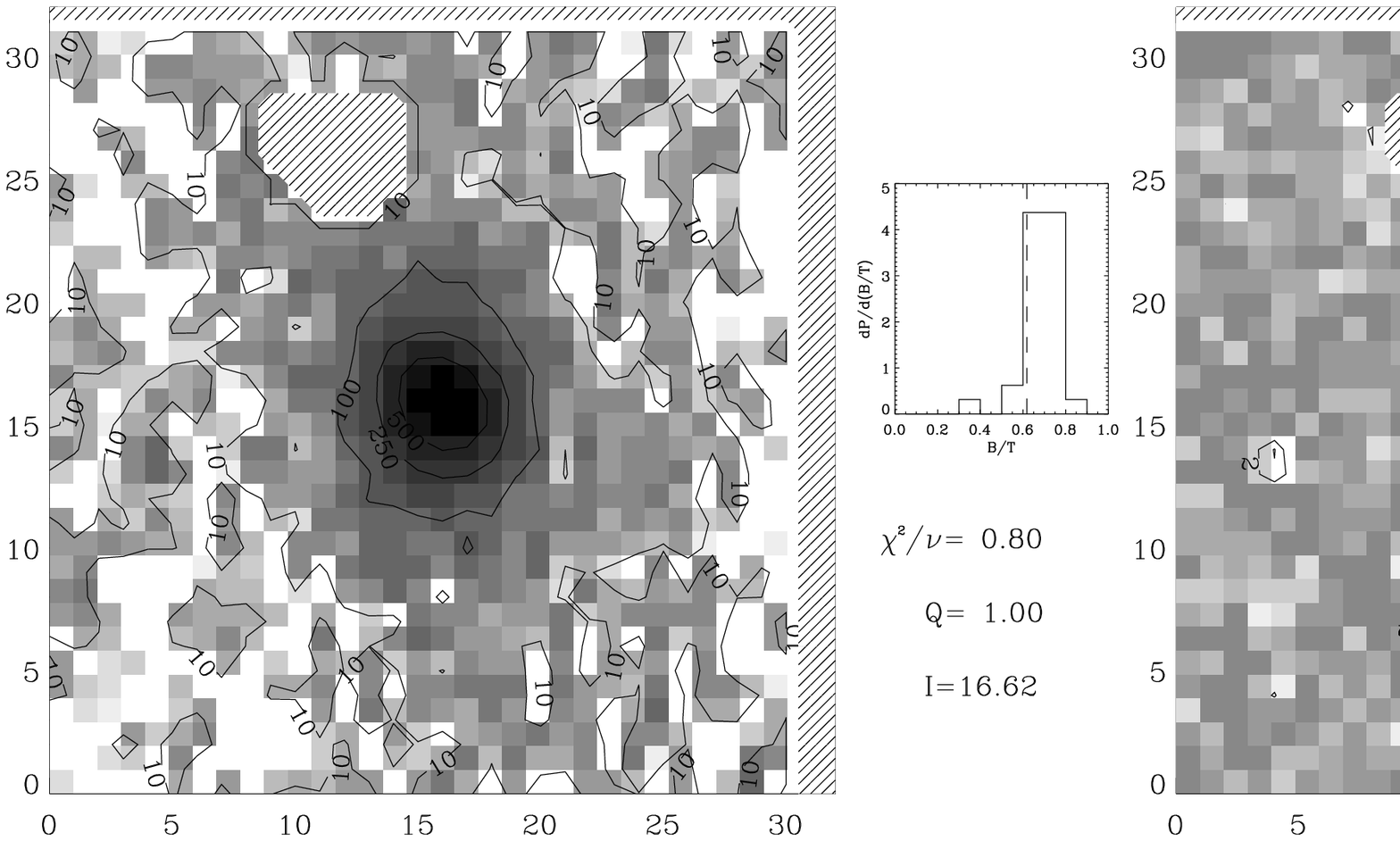} \\
\plotone{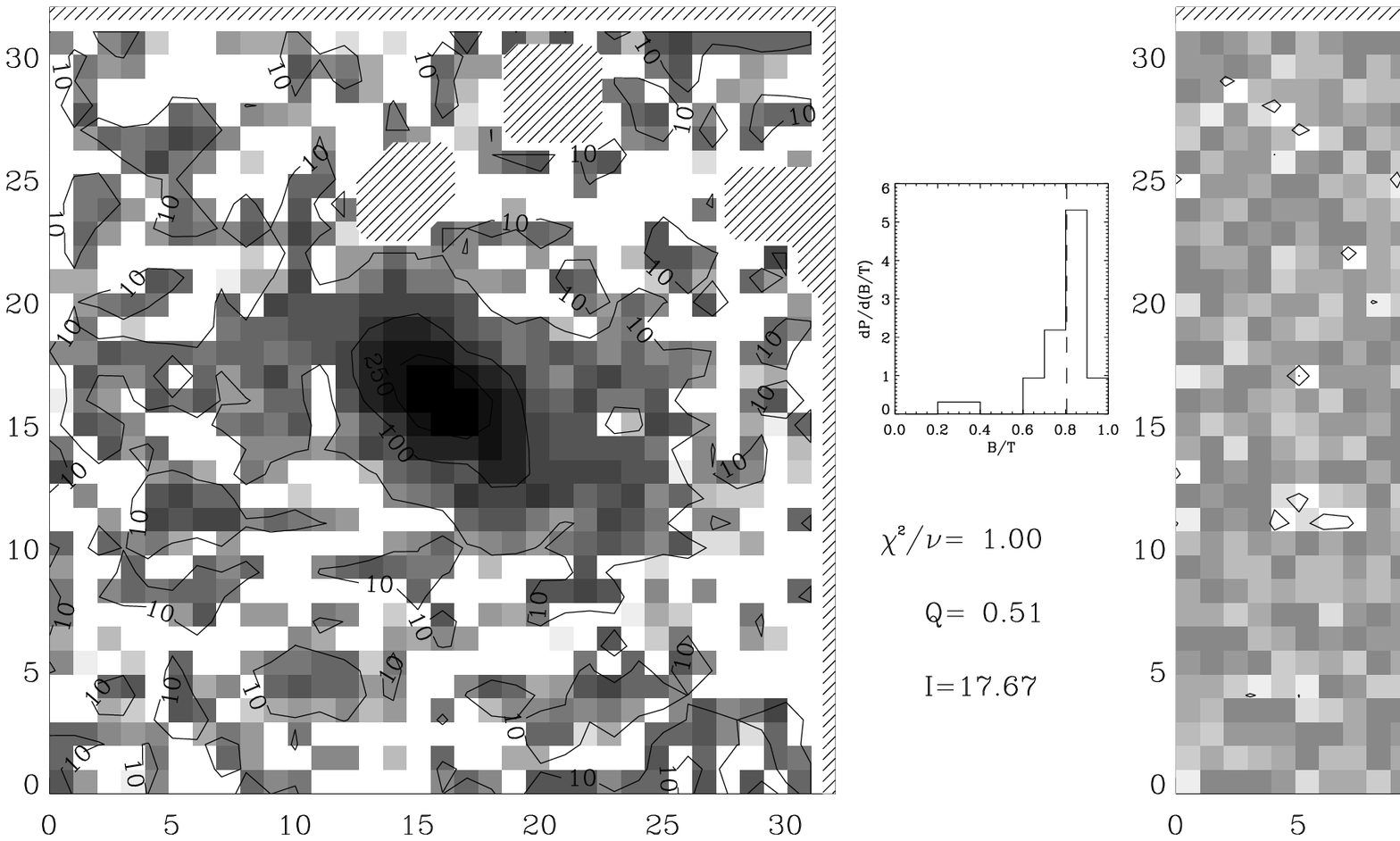} \\
\plotone{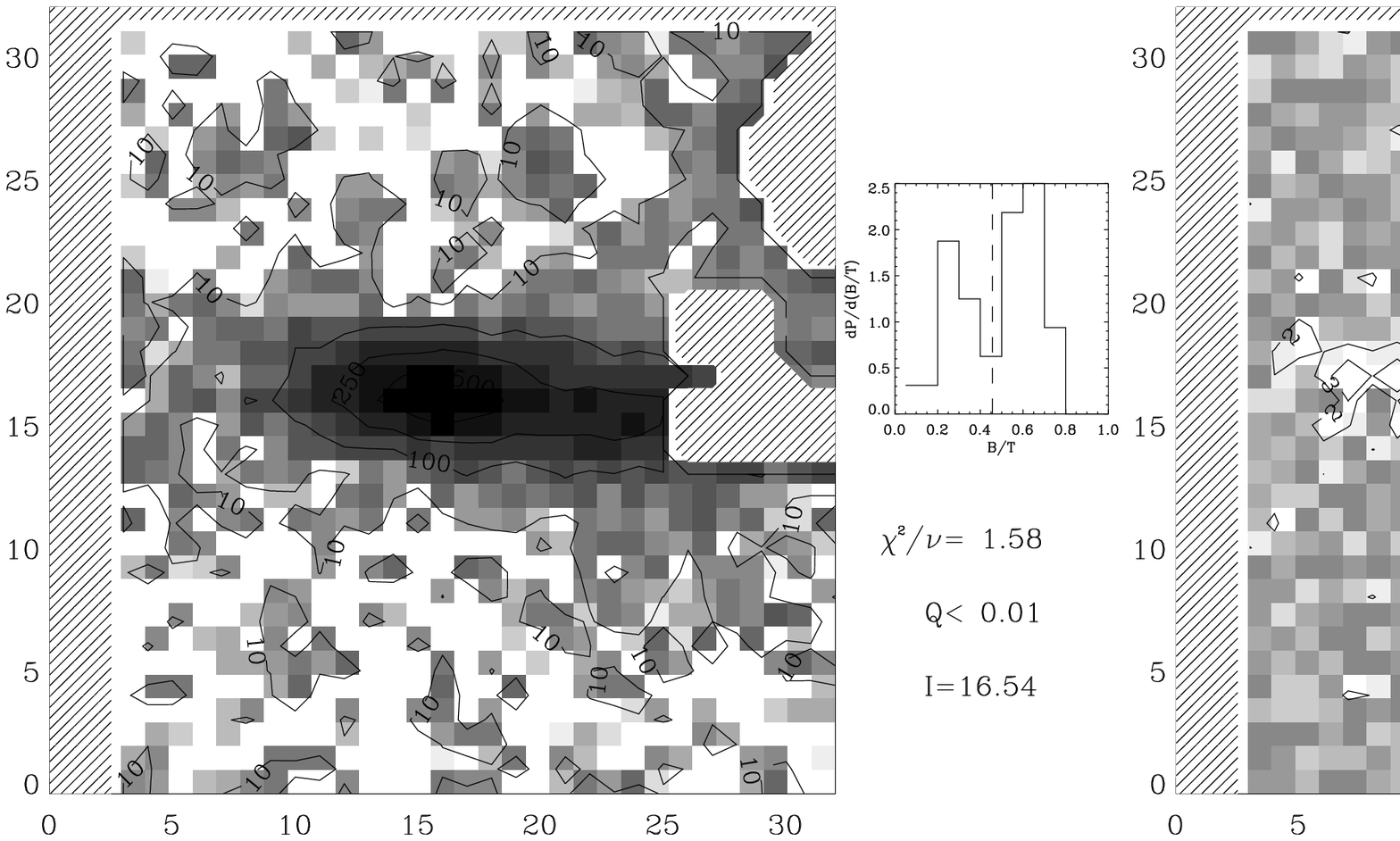}
\caption{\emph{(cont.)} Postage stamp images of three representative
galaxies from our sample. The top row shows a galaxy that is fit very
well by our procedure, the bottom row shows one that is poorly fit and
the middle row shows a more typical result. The left hand column is
the original $33 \times 33$ pixel galaxy image with contours
indicating the pixel value in ADUs. The right hand column is the
residual image after subtracting the best fit model galaxy.  Contours
show the absolute value of the residual in units of $\sigma$, the rms
uncertainty on each pixel value. Hatched regions contained contaminant
galaxies and were removed by our automated cleaning procedure before
fitting. (Where an entire row or column is hatched, the postage stamp
image was recentered prior to fitting.) Between the original image and
the residual maps we quote the value of $\chi^2$ per degree of
freedom, $Q$ (the probability that a random fluctuation exceeds this
value of $\chi^2$) and the I-band apparent magnitude. Also shown is a
histogram of ${\rm d}P/{\rm d}({\rm B/T})$, the distribution of
bulge-to-total ratios found from the Monte Carlo simulations described
in the text, with a vertical dashed line indicating the best-fit B/T
value for the original image.}
\label{fig:resids}
\end{figure*}

\clearpage

The majority of galaxies (363 out of 626) are fit reasonably well by
this procedure. We regard a galaxy as being reasonably well fit when
$Q$, the probability that the measured value of $\chi^2$ is exceeded
by random fluctuations, is greater than 5\%. Not surprisingly though,
many galaxies are not well fit. These typically show signs of strong
morphological disturbance (perhaps due to a recent or imminent merger)
or other inhomogeneities. Examples of well-fit and poorly-fit galaxies
are given in Figure~\ref{fig:resids}. Poorly-fit galaxies are easily
identified by their large $\chi^2$ values and so may be excluded from
further analysis if desired. It should be noted that a poor-fit does
not indicate a failure of our fitting procedure
\emph{per se}, rather it signals that the galaxy is not well described by a
combination of a spheroid and a disk. We choose to show results
computed using the entire sample, regardless of how well a galaxy was
fit, but we will comment on how our results change if badly fit
galaxies are excluded from the analysis.

Errors on the fitted parameters could, in principle, be determined
using a $\Delta \chi^2$ approach, but this would require mapping $\chi
^2$ in the 12 dimensional parameter space of the fit --- an
exceedingly time consuming exercise --- and, in any case, the errors
are unlikely to be normally distributed given that the model is highly
nonlinear in the parameters. We therefore adopt a Monte Carlo approach
to error estimation. Using the best fitting model for each galaxy, we
generate 30 realizations of that model, add random noise at the same
level as in the real image, and mask out any pixels which were masked
out in the original. We then find the best fitting parameters for each
realization and take their distribution as indicative of the
uncertainties in the actual fit. It should be noted that this is only
a valid procedure if the original image is well fit by the model. In
cases where this is not the case, there is no reason to expect the
Monte Carlo distributions to give an estimate of the true errors. The
B/T distributions for the three galaxies illustrated in
Figure~\ref{fig:resids} are shown in that figure.

\section{Results}
\label{sec:results}

\subsection{Accuracy Checks}
\label{sec:acc}

\begin{figure*}
\epsscale{.75}
\plottwo{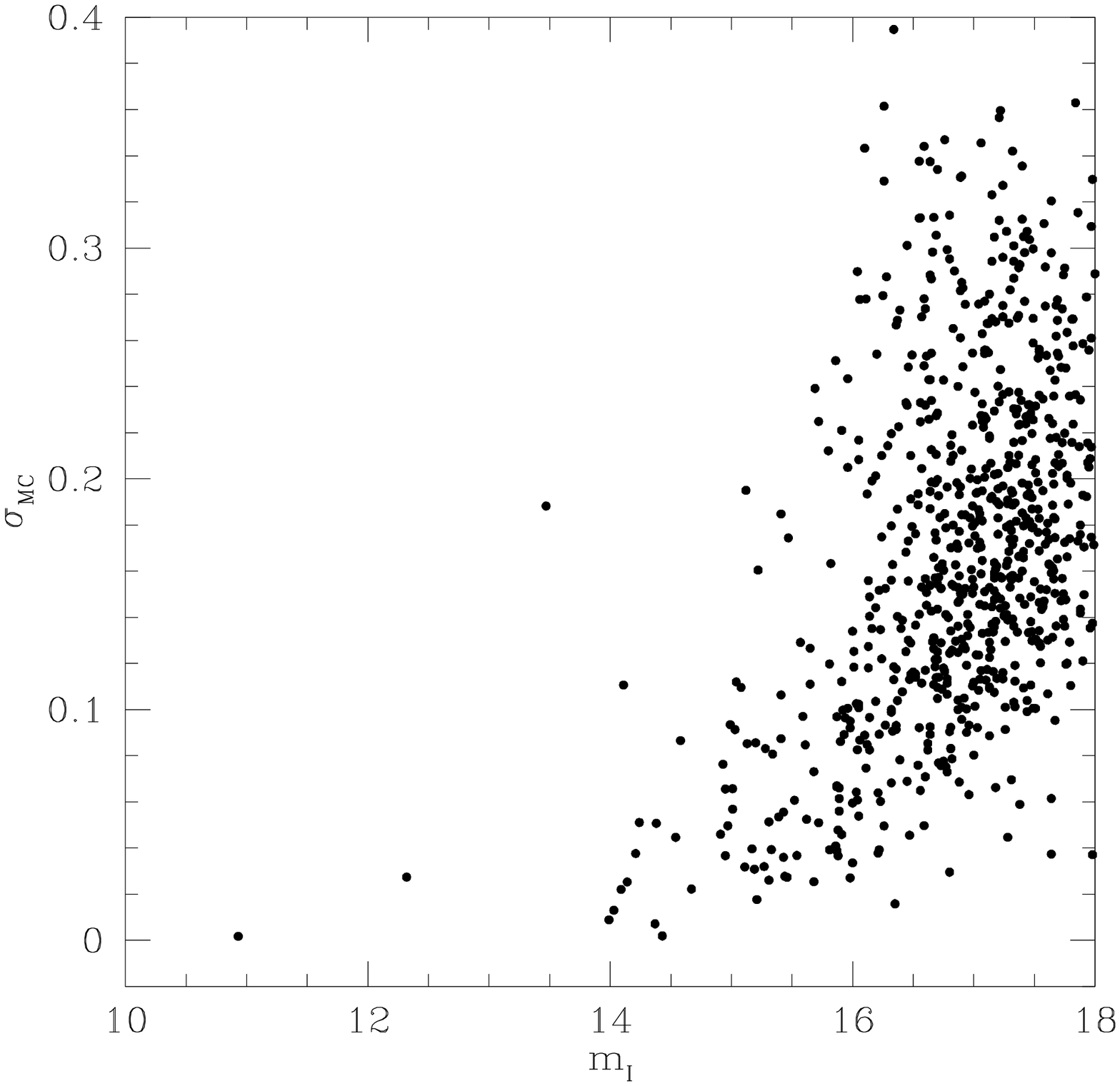}{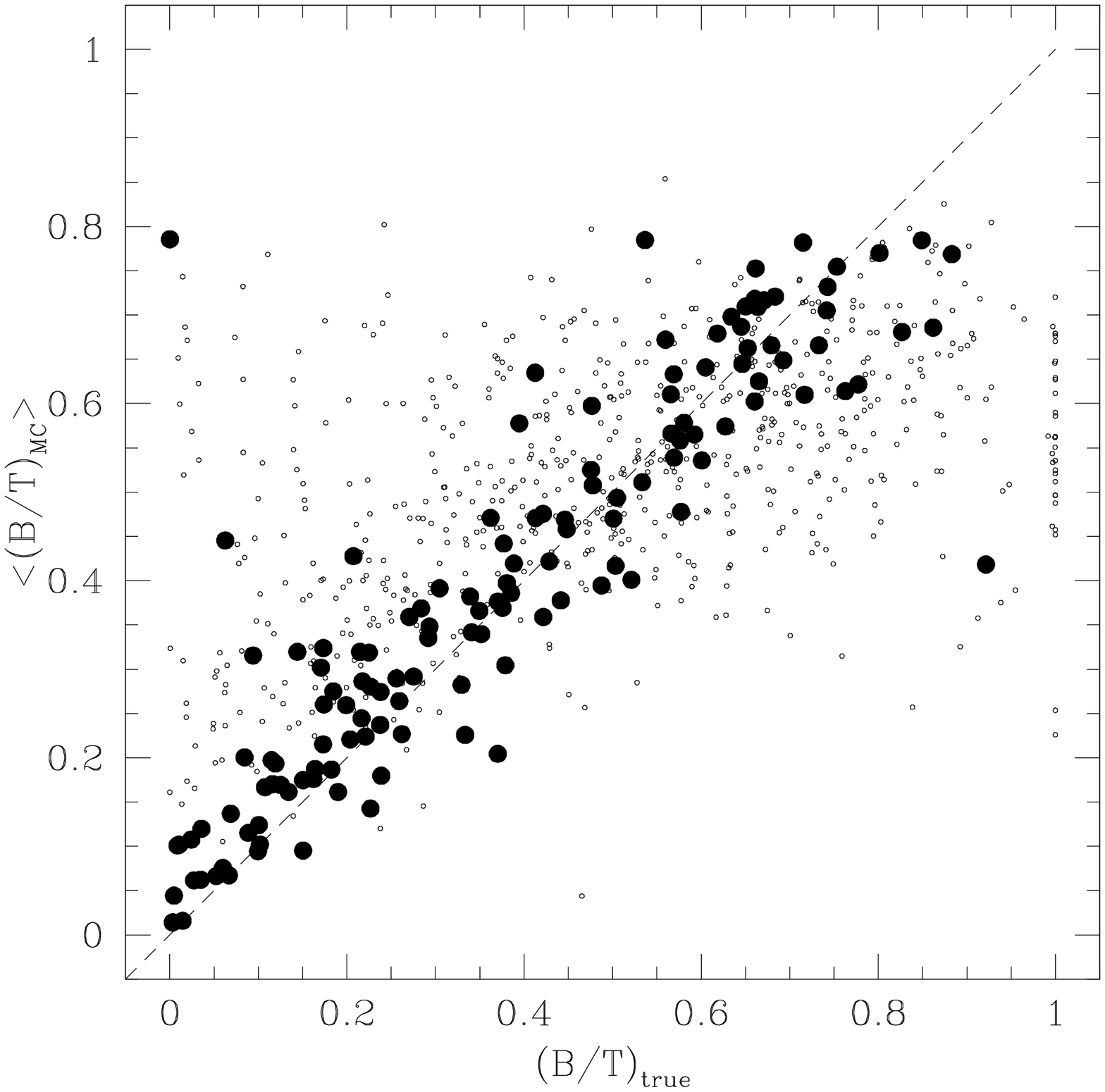}
\caption{\emph{Left-hand panel:} the standard deviation, $\sigma_{\rm
MC}$, of the distribution of recovered B/T ratios from 30 Monte Carlo
realizations of a model galaxy image, as a function of the apparent
I-band magnitude of the model image. \emph{Right-hand panel:} the mean
recovered B/T ratio from 30 Monte Carlo realizations of a model galaxy
image plotted against the true B/T value.  Large, filled circles show
those galaxies for which $\sigma_{\rm MC}\leq 0.1$, while small open
circles show all other galaxies.}
\label{fig:accuracy}
\end{figure*}

We begin by assessing the reliability of our procedure for recovering
the true B/T ratio of a galaxy (assuming, of course, that real
galaxies are well described by our model). Our Monte Carlo procedure
for error estimation allows a determination of the accuracy of
our technique. For each galaxy, the value of B/T input into the Monte
Carlo simulations may be compared to the mean and standard deviation
of the distribution of 30 recovered B/T values.
Figure~\ref{fig:accuracy} gives the results of these accuracy
tests. The left-hand panel shows the standard deviation of the
recovered B/T ratio, as a function of the I-band apparent magnitude of
the mock image. For bright galaxies ($m_{\rm I}\lsim 16$),
$\sigma_{\rm MC}$ is fairly small, typically less than about
0.1. However, for fainter galaxies, $\sigma_{\rm MC}$ increases very
rapidly, resulting in rather poorly constrained B/T values. (In
reality, $\sigma_{\rm MC}$ depends also upon the other parameters
that describe the mock image, but the correlation with apparent magnitude
is the most important.)

In the right-hand panel of Figure~\ref{fig:accuracy}, we plot the mean
value of B/T recovered from the Monte Carlo simulations against the
true value for the mock image. The large solid circles indicate those
images for which $\sigma_{\rm MC}\leq 0.1$. Evidently, for these
galaxies the value of B/T is recovered accurately and without any
strong systematic bias. The small dots show the results for all other
galaxies. Now the scatter is much larger and, more importantly, there
are systematic biases in the mean recovered B/T, such that very low
and very high values are avoided. This effect is not surprising: the
values of B/T for these faint galaxies are almost entirely
unconstrained. (Note that the standard deviation for a completely
uniform distribution of B/T is approximately 0.3.) As a result, the
distribution of B/T from the Monte Carlo simulations becomes close to
uniform, with the mean tending towards 0.5 as $\sigma_{\rm MC}$
increases. For the 90 galaxies in our sample with $m_{\rm I}\leq 16$,
there is a tight correlation between $(B/T)_{\rm true}$ and the mean
value recovered from the Monte Carlo simulations and it is this
subsample that we will use below to compute luminosity
functions. Unfortunately, its relatively small size limits the
statistical accuracy of our estimates quite considerably.

\clearpage

\clearpage

To find the best fit solution we must choose initial values for the
parameters to be fit and then use Powell's algorithm to search for
values producing a better fit. For parameters such as the position
angles, disk inclination and spheroid/disk sizes we make initial
guesses based on the image being fitted. Other parameters are
initially assigned ``typical'' values. We have checked the effect of
altering these initial values. For galaxies with $\sigma_{\rm MC}\leq
0.1$, the choice of initial value makes almost no difference to the
recovered values of the parameters, indicating that our technique is
finding the true minimum $\chi^2$. As $\sigma_{\rm MC}$ becomes
larger, however, the recovered parameters begin to depend strongly
upon the initial values chosen. For these images, the $\chi^2$ surface
in the 12 dimensional parameter space does not possess an obvious
minimum (i.e. it is very noisy). This is just another way of saying
that the values of the fitted parameters for these faint images are
highly uncertain.

\cite{APB95} have demonstrated that the bulges of spiral galaxies are
more accurately fit by an $r^{1/n}$, rather than by the more usual
$r^{1/4}$, surface brightness profile, with values of $n$ ranging from
around 1 to 6 (similar variations in $n$ are seen for elliptical
galaxies; \citealt{bingelli91}; \citealt{caon93}). They show that the
value of $n$ is strongly correlated with morphological type. Although
our present data provide only rather poor constraints on the value of
$n$ (due to limited angular resolution and signal-to-noise), we have
nevertheless repeated the fitting procedure using $r^{1/n}$ profiles
for the spheroids, treating $n$ as a free parameter. For galaxies
where the B/T ratio is well determined, we find that there is a strong
correlation between the B/T values obtained with $r^{1/4}$ and
$r^{1/n}$ profiles, although inevitably some scatter is present. The
disk and spheroid luminosity functions computed using B/T ratios from
$r^{1/n}$ fits show no statistically significant difference from those
using $r^{1/4}$ fits.

Finally, we remind the reader that our analysis makes use of a Gaussian
PSF to mimic the effects of seeing in the data. A Gaussian accurately
describes the core of the PSF measured from bright stars in the
images. However, a profile consisting of a Gaussian core plus
power-law wings provides a better match to many of the stellar
profiles. (The variation of Gaussian and core components from night to
night in the imaging data is not so well characterized, however, and
this is why we make use of a simple Gaussian for our main analysis.)
Fitting the images using such a profile (keeping the relative
proportions of Gaussian and power-law wings fixed, but allowing the
overall radial scale of the PSF to be a free parameter) results in
small changes in the B/T ratio, typically significantly smaller than
the error in the best fit value. Thus, the luminosity functions
presented below are unaffected by the exact choice of PSF. However, it
is clear that a good characterization of the PSF and its variation
will be crucial to obtain accurate disk and spheroid luminosity
functions from larger, higher quality datasets.

\subsection{Luminosity Functions}

Using the I$<16$ sample of approximately 90 galaxies for which we have
good estimates of the B/T ratio we now proceed to estimate the disk
and spheroid luminosity functions. Our aim here is to develop the
techniques required for this measurement and demonstrate them using a
particular dataset. Given the small size of the dataset we must expect
that both statistical (due to the small number of galaxies) and
systematic (due, for example, to the lack of rich clusters in the
dataset) errors will be present. These issues are considered further
in \S\ref{sec:disc}.

To determine the present-day luminosity functions, we need to apply
k$+$e corrections to the galaxy luminosities. We use the
type-dependent k$+$e corrections obtained by
\cite{gardner97}. Briefly, a set of model galaxy colors was computed
using an updated version of the \citet{bc97} stellar population models
with a range of star formation histories.  The observed colors of each
galaxy were matched to one of the models and that particular model was
then used to extrapolate the observed galaxy luminosity to $z=0$. Note
that our type-dependent k$+$e corrections are based on the
\emph{total} (i.e. disk plus spheroid) color of each galaxy. In
principle, k$+$e corrections could be applied to each component
separately if spheroid/disk decompositions were carried out in several
bands. Given the uncertainties in our present estimates of B/T, we
refrain from this degree of complexity in this analysis.

We use the stepwise maximum likelihood (SWML) estimator proposed by
\cite[hereafter EEP]{eep} and also the parametric maximum likelihood
method proposed by \cite[hereafter STY]{STY} to compute disk and
spheroid luminosity functions. The detectability of a spheroid depends
on both its apparent magnitude and the B/T ratio, and we must account
for this in constructing the likelihood function. This leads us to
define a two-dimensional function, $\Phi(M,B)$, such that
$\Phi(M,B){\rm d}M{\rm d}B$ is the number of galaxies per unit volume
with B/T ratio in the range $B$ to $B+{\rm d}B$ and \emph{spheroid}
absolute magnitude in the range $M$ to $M+{\rm d}M$ (with an
equivalent definition for disks). The application of the maximum
likelihood estimator to this function is discussed in detail in
Appendix~A. The normal luminosity function of spheroids is readily
derived using $\phi(M)=\int_0^1 \Phi(M,B){\rm d}B$ (and similarly for
disks). For the STY method we must assume some parametric form for the
luminosity function. We have tried fitting the disk and spheroid
luminosity functions with a ``Schechter$\otimes$exponential'' form,
namely $\Phi(M,B)=\phi(M) \exp(\beta B)$, where $\phi(M)$ is the
normal Schechter function and $\beta$ is a parameter to be fit,
motivated by the shape of the SWML estimate of these luminosity
functions.

\clearpage

\begin{figure}
\plotone{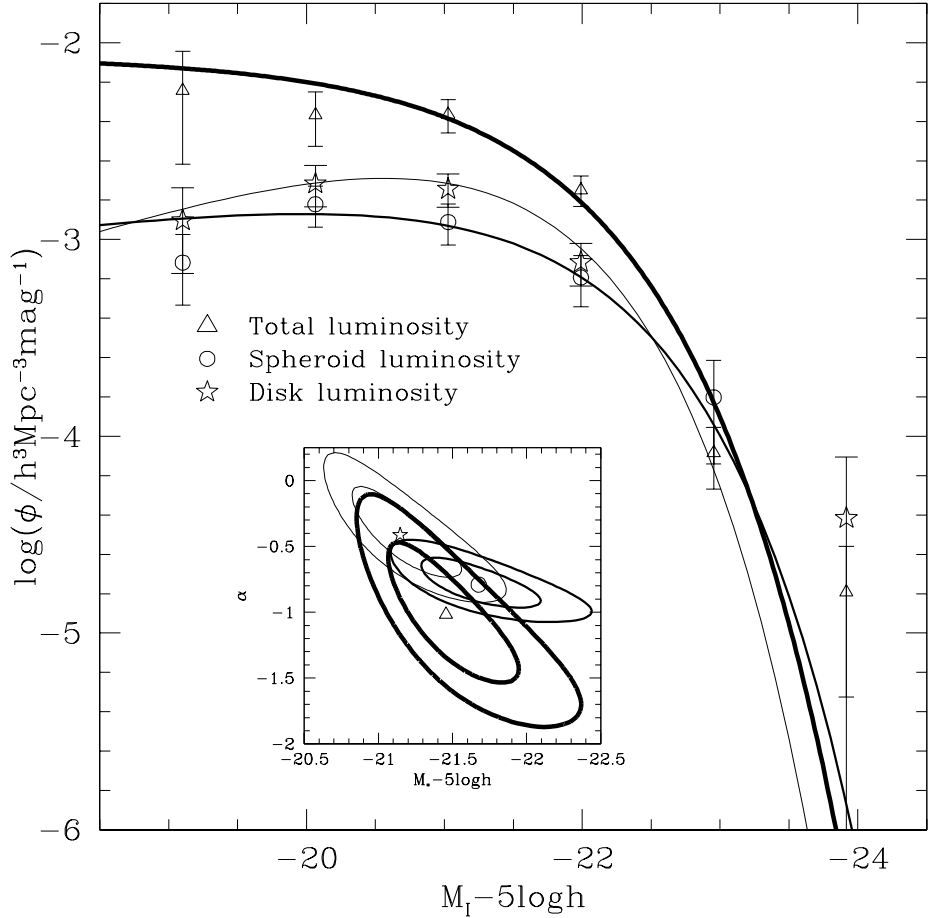}
\caption{I-band luminosity functions. Triangles, circles and stars
show the SWML estimates of the total, spheroid and disk luminosity
functions respectively. Only galaxies brighter than $I=16$ have been
used, k$+$e corrections have been applied to all galaxies, and
distances have been calculated assuming
$(\Omega_0,\Lambda_0)=(0.3,0.7)$. Errorbars are the sum in quadrature
of the standard SWML errors and the variance in estimates of the
luminosity function from 30 Monte Carlo realizations of the
spheroid/disk decomposition procedure. The very heavy solid line shows
the best fit Schechter function to the total luminosity function,
while the inset shows the values of $\alpha$ and $M_\star$ for this
fit, together with their 1 and $2\sigma$ error ellipses. The heavy and
thin solid lines show the best-fit ``Schechter$\otimes$exponential'' functions
to the spheroid and disk luminosity functions respectively, and
confidence regions for $\alpha$ and $M_\star$ for these fits are given
in the inset (the remaining parameter of the fits was $\beta=0.0\pm
0.37$ and $2.1\pm 0.37$ for spheroids and disks respectively).}
\label{fig:ILF}
\end{figure}

\clearpage

Figure~\ref{fig:ILF} shows the resulting I-band luminosity functions
with distances computed assuming
$(\Omega_0,\Lambda_0)=(0.3,0.7)$\footnote{Assuming
$(\Omega_0,\Lambda_0)=(1,0)$ instead changes our results only
slightly, shifting the luminosity function faintwards due to the
smaller luminosity distance in this model.} and $H_0=100
h$\,km/s/Mpc. Triangles show the total luminosity function; circles
and stars show the spheroid and disk luminosity functions
separately. These SWML luminosity functions are normalized to the
I-band number counts in the Sloan Digital Sky Survey using the
procedure described in the Appendix. For the total luminosity
function, we plot the standard SWML errorbars (obtained from the
covariance matrix of the luminosity function as described by EEP), but
for the spheroid and disk luminosity functions, the errorbars are the
sum in quadrature of the standard SWML errors and the variance in the
luminosity function estimated from the 30 Monte Carlo realizations of
the spheroid/disk decomposition process. The errors from each source
are of comparable magnitude (although the variance from the Monte
Carlo realizations is the smaller of the two). The very heavy solid
line shows the best-fitting Schechter function to the total luminosity
function (determined using the STY method); the inset shows the values
of $\alpha$ and $M_\star$ for this fit, together with their 1 and
$2\sigma$ error contours. (The small sample size is reflected in
rather large and correlated uncertainties in $M_\star$ and $\alpha$.)
Heavy and thin solid lines show the best fit STY
``Schechter$\otimes$exponential'' luminosity function fits to the spheroid and
disk luminosity functions respectively (with the corresponding
confidence ellipses for $\alpha$ and $M_\star$ shown in the inset, and
the values of $\beta$ given in the figure caption). A likelihood ratio
test \citep{eep} shows that the Schechter$\otimes$exponential luminosity
function is not a particularly good fit to the data. With the present
small dataset we have been unable to find a better functional
form. This situation will be rectified with a larger dataset (assuming
that some suitable functional form does actually exist).

The I-band luminosity functions of disks and spheroids are remarkably
similar. The only significant difference is that the spheroid
luminosity function is somewhat lower at faint magnitudes. However,
given the small size of the present sample, this difference may not be
robust. The luminosity densities in disks and spheroids obtained by
integrating the SWML luminosity functions over the range of absolute
magnitudes shown in Fig.~\ref{fig:ILF} are $5.8 \pm 0.8$ and $4.7 \pm
0.7 \times 10^7 h L_\odot/\hbox{Mpc}^3$ respectively (where we have
taken $M_\odot=4.14$ in the I-band; \citealt{cox00}). In principle, we
can use our Schechter$\otimes$exponential fits to estimate the total
luminosity density, extrapolating to include the contribution from
arbitrarily faint spheroids and disks. Doing so yields results which
agree with the SWML estimates within the quoted errors, suggesting
that our determination may have suitably converged. However, it must
be kept in mind that the Schechter$\otimes$exponential form is not a
particularly good fit to the current datasets.

\clearpage

\begin{figure}
\plotone{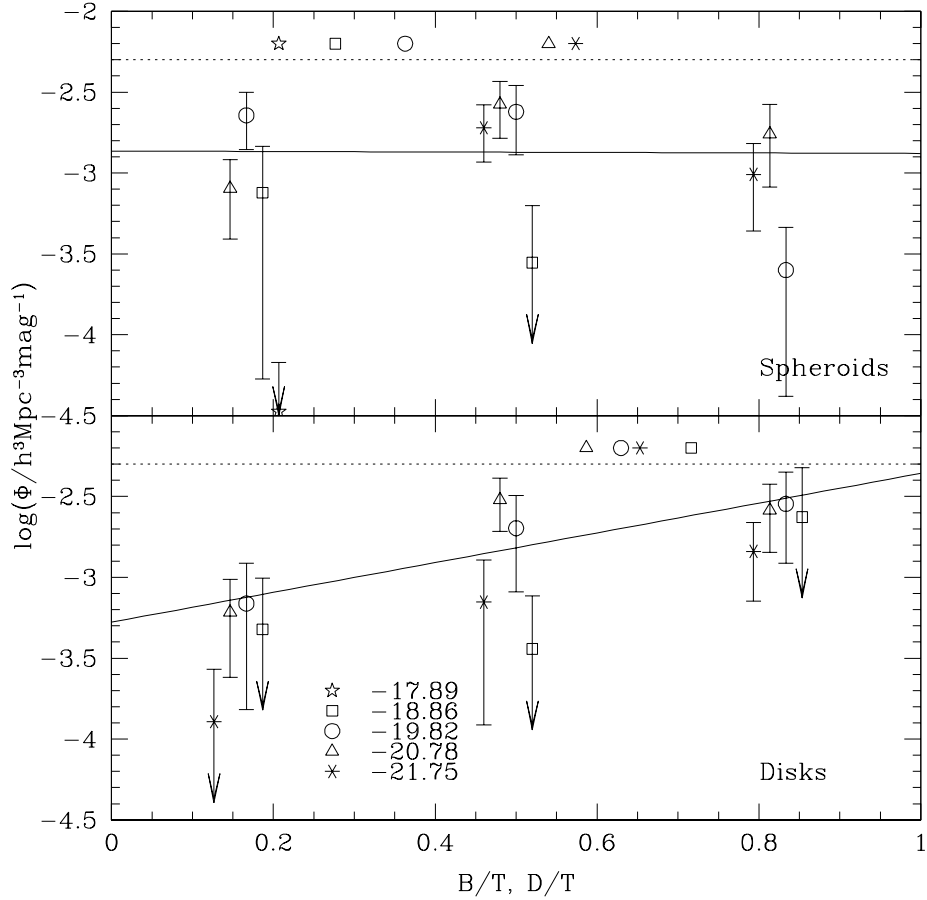}
\caption{Slices through the SWML estimate of the bi-variate luminosity
function, $\Phi(M,B)$, for different absolute magnitudes, $M$, as
indicated in the figure. Points were computed in bins of size $(\Delta
M,\Delta B)=(0.48,0.33)$ and errors obtained as described by EEP. The
upper panel shows the spheroid luminosity function, and the lower
panel the disk luminosity function. The points without errorbars above
the dotted line in each panel indicate the mean B/T (and D/T) for
spheroids (and disks) in the corresponding absolute magnitude
bins. Solid lines indicate the best-fit ``Schechter$\otimes$exponential''
parametric luminosity function for the $M_{\rm I}-5\log h=-19.82$
bin.}
\label{fig:rat}
\end{figure}

\clearpage

\begin{figure}
\plotone{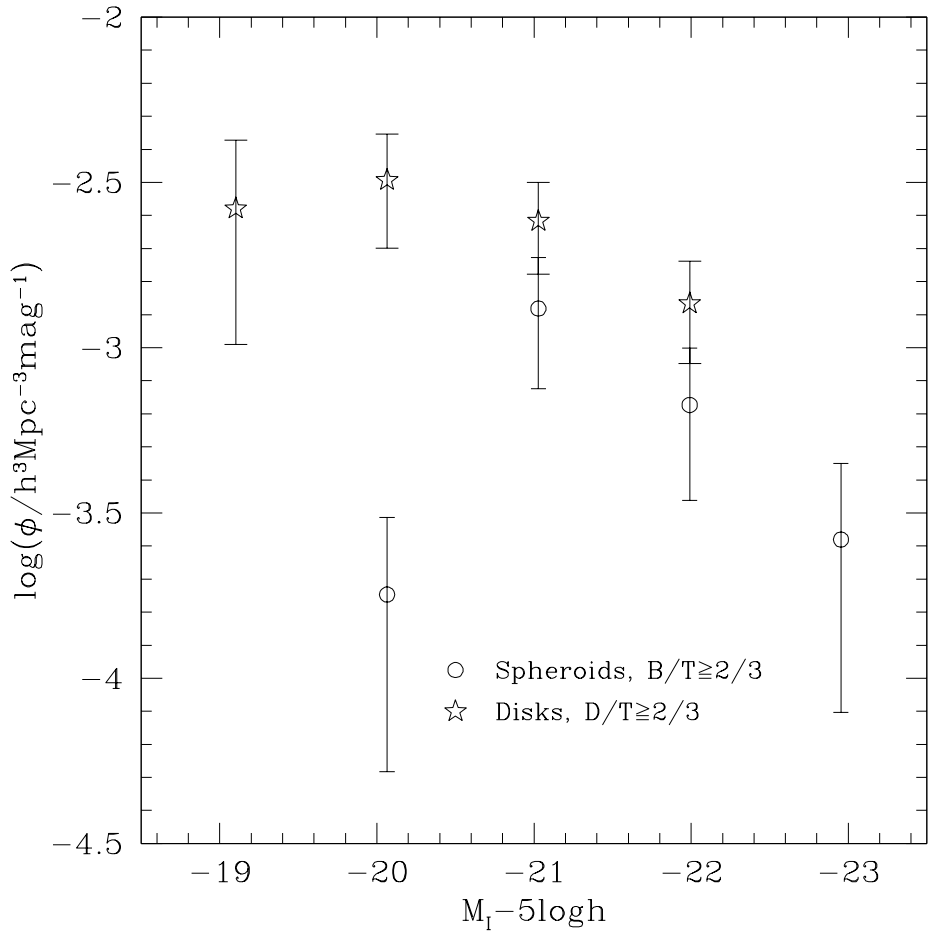}
\caption{SWML estimates of the I-band luminosity functions of
spheroids in spheroid-dominated galaxies (i.e. B$>2/3$; circles) and
disks in disk-dominated galaxies (i.e. D/T$>2/3$; stars). The samples
include 13 and 36 galaxies respectively. Errors are the sum in
quadrature of the standard SWML errors and the variance found in
luminosity functions estimated from 30 Monte Carlo realizations of the
spheroid/disk decomposition procedure.}
\label{fig:EvsS}
\end{figure}

In Fig.~\ref{fig:rat}, we show slices through the bi-variate
luminosity function, $\Phi(M,B)$, at constant $M$ for several values
of $M$ (as indicated in the figure, and in bins of width $\Delta
M=0.48$). It is evident that $\Phi(M,B)$ is \emph{not} independent of
$B$ and that, in fact, it may not be separable into a simpler form
$\Phi(M,B)=\phi(M) g(B)$. This is particularly noticeable for the
spheroid luminosity function.  Figure~\ref{fig:rat} shows that
spheroid-dominated systems (i.e. B/T$>2/3$) are found only in the
brightest spheroids, while disk-dominated systems (i.e. D/T$>2/3$)
have disks with a much broader range of luminosities. This point is
made more clearly in Fig.~\ref{fig:EvsS} where we show the spheroid
luminosity function of spheroid-dominated systems and the disk
luminosity function of disk-dominated systems. We find
spheroid-dominated systems in abundance only brightwards of $M_{\rm
I}-5\log h\approx -21$, but disk-dominated systems across the whole
range of luminosities.

\clearpage

\clearpage

\subsection{The stellar mass in disks and spheroids}

The stellar mass associated with the luminosity of each galaxy is
easily obtained by a similar procedure to that employed to calculate
the k-corrections (namely fitting their BVIK colors to a set of
template galaxies).  We use these estimates to construct the stellar
mass functions of spheroids and disks in the local Universe and
thereby estimate the total mass content in each component. The method
that we adopt is the same that \cite{cole01} used to measure the {\em
total} stellar mass density in the Universe from a combination of
2dFGRS and 2MASS data. \cite{cole01} found the stellar mass
density\footnote{Specifically, \protect\cite{cole01} estimated the
mass locked up in stars and stellar remnants which differs from the
time integral of the star formation rate due to recycling of material
by massive stars. We adopt the same definition of stellar mass here.}
in units of the critical density to be $\Omega_{\rm
stars}=0.0016\pm0.00024h^{-1}$ or $\Omega_{\rm
stars}=0.0029\pm0.00043h^{-1}$, depending on whether a \cite{k83} or
a \cite{sal} stellar initial mass function (IMF) was assumed.  (Both
estimates include the effects of dust on galaxy luminosities as
described by \citealt{cole01}.)  We normalize our stellar mass functions
by requiring them to produce the same number density of galaxies more
massive than $M_*$ as the \cite{cole01} stellar mass function.  The
total stellar mass density inferred from our ${\rm I}<18$ sample is
then $\Omega_{\rm stars}=0.0009\pm0.00007h^{-1}$ or $\Omega_{\rm
stars}=0.0017\pm0.00012h^{-1}$ for the same two IMFs respectively and the same
prescription for dust-extinction.  Errors on the stellar mass density
were found by summing in quadrature the error from each individual bin
in the SWML mass function, together with the error in the overall
normalization.  Our estimates include contributions from galaxies with
stellar masses greater than $10^9h^{-2}M_\odot$ below which the SWML
stellar mass function is not well determined. We can check our result
using the STY stellar mass function. For this, we fit a Schechter
function convolved with a Gaussian of width $0.1$ in $\log_{10}M_{\rm
stars}$ to account for the scatter in the relation between stellar
mass and I-band absolute magnitude. We find that our estimates of
$\Omega_{\rm stars}$ using the SWML and STY mass functions agree
within the errors, suggesting that the result has already converged to
sufficient accuracy.  Our estimates, however, are lower than those of
\cite{cole01}, a reflection of the flatter faint end slope of our
mass functions which may well be due to the small size of our sample.

For our ${\rm I}<16$ sample, for which the bulge-to-total ratio is
well measured, we find a slightly higher total stellar mass density of
$\Omega_{\rm stars}=0.0012\pm0.00014h^{-1}$ for the \cite{k83}
IMF. Splitting into spheroidal and disk components, we find
$\Omega_{\rm stars,spheroids}=0.00039\pm0.00006h^{-1}$ and
$\Omega_{\rm stars,disks}=0.00051\pm0.00008h^{-1}$ for this same
IMF. (Note that with the SWML method the stellar mass densities of
disks and spheroids are not guaranteed to sum to give the total
stellar mass density.)  If, instead, we assume a Salpeter IMF, the
ratio of disk to spheroid stellar mass densities increases slightly
from 1.31 to 1.37, but this change is negligible given the current
errors in these quantities. Although the small size of our sample is
clearly a significant limitation, this initial result suggests that
spheroids and disks contribute about equally to the stellar mass
density of the Universe. The techniques developed in this paper, when
applied to a much larger dataset, should allow their contributions to
be more accurately determined.

\section{Discussion}
\label{sec:disc}

We have presented a detailed method to determine the bulge-to-total
ratios of galaxies by fitting to two-dimensional photometry, and have
applied this technique to determine the I-band bulge-to-total
luminosity ratios of a sample of approximately 600 galaxies brighter
than I$=18$ with spectroscopic redshifts. Our approach is designed to
work with realistic galaxy images, dealing automatically with
contamination by nearby objects, a varying PSF and small changes in
the background from image to image. A crucial part of the fitting
procedure is a Monte Carlo determination of the errors on the fitted
parameters, an approach which is favored since it is fast and
automatically accounts for the highly non-linear nature of the model
parameters. For the current sample of galaxies, around 60\% are well
fit by a combination of an exponential disk and an $r^{1/4}$-law
spheroid. Those that are not well fit frequently show signs of
morphological disturbance. We find that bulge-to-total ratios are
determined accurately (i.e. with errors of around $10\%$) only for
galaxies brighter than I$\approx 16$.

For the 90 galaxies brighter than I$=16$ in this sample we measure the
B/T ratio with reasonable accuracy. We have used the resulting
disk/spheroid decomposition of these bright galaxies to construct
separate luminosity functions for disks and spheroids. We find no
significant differences between them when considered purely in terms
of luminosity, although the statistical uncertainties associated with
the small sample size make the detection of any differences
difficult. However, when we consider the bi-variate distributions of
luminosity and bulge-to-total or disk-to-total light, we find that
spheroid dominated systems (B/T$>2/3$) only occur for the brightest
spheroids, while disk-dominated systems (D/T$>2/3$) occur for a much
broader range of disk luminosities.

The relative contributions of disks and spheroids to the total stellar
mass density in the Universe is a very important constraint on
theories of galaxy formation which attempt to describe the assembly of
galaxies as a function of time. We find, perhaps surprisingly, that
the disks and spheroids in our sample contribute almost equally to the
stellar mass density today (in a ratio of $1.3 \pm 0.2$). Since the
stellar populations in disks are generally younger than those in
spheroids, it is an interesting coincidence that the total stellar
mass in the two kinds of structural components should be so similar at
the present time. 

\citet{schecht87} reached a similar conclusion to ours 
using a photometric comparison technique to estimate the B-band
bulge-to-disk ratios of galaxies. While this technique may not be as
accurate as our own on a galaxy-by-galaxy basis, it should provide a good
estimate of the total contribution of each component to the stellar mass
density. It is therefore reassuring that our results agree well with those
of \citet{schecht87}.  A different result was obtained by \citet{fuku98},
who derived a ratio of disk to spheroid stellar mass density of $0.33 \pm
0.23$. Although they found comparable B-band luminosity density in
spheroids and disks, they adopted a spheroid mass-to-light ratio around
four times greater than that for disks, resulting in spheroids making a
significantly greater contribution to the stellar mass density. While we
use a more accurate technique for converting from luminosity to stellar
mass (a technique which could be improved further if B/T ratios were
measured for each galaxy in several bands), the small size of our sample
limits the accuracy of our results. In particular, our sample may not
contain enough rich clusters which are known to contain higher fractions of
spheroid dominated galaxies than the field (e.g. \citealt{dressler80}) and
this could introduce a small bias in our results.

Clearly the greatest limitation of this work is the small size of the
sample of galaxies for which accurate disk/spheroid decompositions can
be performed. Fortunately, this problem should be remedied in the near
future with the advent of high quality, large area photometric surveys
such as that being carried out by the SDSS project.

\section*{Acknowledgments}

We thank Jon Gardner and Carlton Baugh for supplying data used in this
work and for valuable discussions, and the referee, Alan Dressler, for
valuable suggestions. We also thank Istvan Szapudi for his assistance
in the early stages of this work.

\section*{Appendix A. Estimators for Spheroid and Disk Luminosity Functions}
\label{sec:estimators}

The traditional $1/V_{\rm max}$ estimator is trivially adapted to the
case of disk and spheroid luminosity functions. The estimator is
applied just as in the case of the standard luminosity function,
except that the \emph{total} luminosity of the galaxy (i.e. disk plus
spheroid luminosity) is used to compute $V_{\rm max}$, since it is
this total luminosity that determines the volume within which the
galaxy could have been detected.

The maximum likelihood estimator of \citet[hereafter EEP]{eep} is also
easily generalized to the case of spheroid and disk luminosity
functions. Consider the case of the spheroid luminosity function (the
same arguments apply to disks). As noted earlier, the detectability of
a spheroid depends upon both its absolute magnitude, $M$, and on the
bulge-to-total ratio which we denote by $B$ in this Appendix. We begin
therefore by defining a two-dimensional function, $\Phi(M,B)$, such
that $\Phi(M,B){\rm d}M{\rm d}B$ is the number of galaxies with
bulge-to-total ratio in the range $B$ to $B+{\rm d}B$ and
\emph{spheroid} absolute magnitude $M$ to $M+{\rm d}M$ per unit
volume. The normal luminosity function of spheroids is easily
recovered using $\phi(M)=\int_0^1 \Phi(M,B){\rm d}B$. The probability
that galaxy $i$ with spheroid magnitude $M_i$ and bulge-to-total ratio
$B_i$ is seen in a magnitude limited survey is
\begin{equation}
p_i \propto \Phi(M_i,B_i) \left/ \int_0^1 \int_{-\infty}^{M^\prime_{\rm lim}(z_i,B)} \Phi(M,B) {\rm d}M {\rm d}B \right. ,
\end{equation}
where $M^\prime_{\rm lim}(z_i,B)=M_{\rm lim}(z_i)-2.5 \log_{10}B$ and
$M_{\rm lim}(z_i)$ is the limiting absolute magnitude of the survey at
redshift $z_i$. The use of $M^\prime_{\rm lim}$ is necessary since
arbitrarily faint spheroids will make it into the survey provided that 
they have a sufficiently low bulge-to-total ratio (corresponding to
sufficiently bright disks).

From this definition we can construct the usual likelihood function
\begin{equation}
\ln {\mathcal L} = \sum_{i=1}^N \ln \Phi(M_i,B_i) - \sum_{i=1}^N \ln \left\{ \int_0^1 \int_{-\infty}^{M^\prime_{\rm lim}(z_i,B)} \Phi(M,B) {\rm d}M {\rm d}B \right\} + \hbox{const},
\label{eq:lnL}
\end{equation}
where $N$ is the total number of galaxies. There are now two ways to
proceed. In the first we assume a simple parametric form for
$\Phi(M,B)$ and maximize the likelihood with respect to the
parameters. This is analogous to fitting a Schechter function
\citep{schecht76} to the normal luminosity function
(e.g. \citealt{STY}). A simple parametric form which we have tried to
fit our data is
\begin{equation}
\Phi(M,B)=\phi(M) \exp(\beta B),
\end{equation}
where $\phi(M)$ is the usual Schechter function and $\beta$ is a
parameter to be estimated from the fit. With this method, the
likelihood function of eqn.~(\ref{eq:lnL}) can be evaluated for each
value of the three parameters $\alpha$, $\beta$ and $M_\star$, and
hence the parameter values which maximize the likelihood are readily obtained.

The second approach involves splitting $\Phi(M,B)$ into bins in $M$
and $B$ and treating each as a parameter. This is equivalent to the SWML
method of EEP for estimating the standard luminosity function.

We represent $\Phi(M,B)$ as follows:
\begin{equation}
\Phi(M,B) = \Phi_{k,h}, \left\{ \begin{array}{ll} M_k-\Delta M/2 < M < M_k + \Delta M/2, & k=1,\ldots,N_p \\  B_h-\Delta B/2 < B < B_h + \Delta B/2, & h=1,\ldots,N_q. \end{array} \right.
\end{equation}
The likelihood function may then be written as
\begin{equation}
\ln {\mathcal L} = \sum_{i=1}^N W(M_i-M_k,B_i-B_h)\ln \Phi_{k,h} - \sum_{i=1}^N \ln \left\{ \sum_{h=1}^{N_q} \sum_{k=1}^{N_p} \Phi_{k,h} \Delta M \Delta B H[M_k,B_h,M_{\rm lim}(z_i)] \right\} + \hbox{const},
\end{equation}
where
\begin{equation}
W_{k,h}(M_i,B_i)=\left\{ \begin{array}{ll} 1 & \hbox{if } M_k-\Delta M/2 < M_i < M_k + \Delta M/2, \hbox{ and } B_h-\Delta B/2 < B_i < B_h + \Delta B/2, \\ 0 & \hbox{otherwise}  \end{array} \right.
\end{equation}
and
\begin{equation}
H[M_k,B_h,M_{\rm lim}(z_i)] = {1 \over \Delta M \Delta B} \int_{B_h-\Delta B/2}^{B_h+\Delta B/2} \int_{M_k-\Delta M/2}^{M_k+\Delta M/2} Q(M,B) {\rm d}M {\rm d}B,
\end{equation}
where $Q(M,B)=0$ if $M>M_{\rm lim}(z_i)-2.5\log_{10}B$ and $Q(M,B)=1$
otherwise. Since only the shape of the luminosity function is
constrained by the above likelihood function, we introduce an
additional constraint, $g=\sum_k \sum_h \Phi_{k,h}(L_{k,h}/L_{\rm
f})^\beta \Delta M \Delta B-1=0$, where $L_{k,h}$ is the total
luminosity of a galaxy with spheroid magnitude $M_k$ and
bulge-to-total ratio $B_h$ and $L_{\rm f}$ is a fiducial luminosity
(which we will take to be that corresponding to $M_{\rm I}-5\log
h=-20.5$), using a Lagrangian multiplier $\lambda$ as did
EEP. Maximizing $\ln {\mathcal L}^\prime = \ln {\mathcal L}+\lambda g$
then yields
\begin{equation}
\Phi_{k,h} = {\sum_{i=1}^N W(M_i-M_k,B_i-B_k) \over \sum_{i=1}^N H[M_k,B_h,M_{\rm lim}(z_i)]/\sum_{l=1}^{N_p} \sum_{m=1}^{N_q} \Phi_{l,m} H[M_l,B_m,M_{\rm lim}(z_i)]}
\end{equation}
which are easily solved with an iterative procedure. The covariance
matrix for the parameters is obtained in a manner entirely analogous
to that outlined by EEP.

Normalization of the maximum likelihood luminosity function can be
achieved using the actual redshift data as described by
\cite{loveday92}, but using $M^\prime_{\rm lim}$ in the selection
function to account for the effects of the bulge-to-total ratio. A
better approach is to normalize by performing a least squares fit to
the number counts of galaxies from a wide area survey. The cumulative
number count to apparent magnitude $m$ is given by
\begin{equation}
n(m) = \int_0^\infty \int_0^1 \int_{-\infty}^{M^\prime_{\rm
lim}}\Phi(m-D(z)-K(z)-2.5\log_{10}B,B) {{\rm d}V \over {\rm d}z} {\rm
d}M {\rm d}B {\rm d}z,
\end{equation}
where $D(z)$ and $K(z)$ are the distance modulus and k$+$e correction
respectively at redshift $z$, which we can compute from the SWML
estimate of $\Phi$.


\begin{thebibliography}{}
\bibitem[Abraham et al.(1996)]{abraham96}Abraham~R.~G., van den Bergh~S., Glazebrook~K., Ellis~R.~S., Santiago~B.~X., Surma~P., Griffiths~R.~E., 1996, ApJS, 107, 1
\bibitem[Andredakis, Peletier \& Balcells(1995)]{APB95}Andredakis~Y.~C., Peletier~R.~F., Balells~M., 1995, MNRAS, 275, 874
\bibitem[Arnaud \& Rothenflug(1985)]{arnaud85}Arnaud~M., Rothenflug~R., 1985, A\&AS, 60, 425
\bibitem[Barnes \& Hernquist(1992)]{barnes92}Barnes~J.~E., Hernquist~L., 1992, ARA\&A, 30, 705
\bibitem[Baugh, Cole \& Frenk(1996a,b)]{baugh96ab}
\vspace{-10mm}\bibitem[Baugh, Cole \& Frenk(1996a)]{baugh96a}Baugh~C.~M., Cole~S., Frenk~C.~S., 1996a, MNRAS, 282, L27
\bibitem[Baugh, Cole \& Frenk(1996b)]{baugh96b}Baugh~C.~M., Cole~S., Frenk~C.~S., 1996b, MNRAS, 283, 1361
\bibitem[Bertin \& Arnouts(1996)]{sex}Bertin~E., Arnouts~S., 1996, A\&AS, 117, 393
\bibitem[Bingelli \& Cameron(1991)]{bingelli91}Bingelli~B., Cameron~L.~M., 1991, A\&A, 252, 27
\bibitem[Brent(1973)]{brent73}Brent~R.~P., 1973, Algorithms for minimization without derivatives (Englewood Cliffs, New Jersey: Prentice Hall), Chapter~7
\bibitem[Bruzual \& Charlot(1993)]{bc97}Bruzual \& Charlot, 1993, ApJ, 405, 538
\bibitem[Byun \& Freeman(1995)]{byun95}Byun~Y.~I., Freeman~K.~C., 1995, ApJ, 448, 563
\bibitem[Caon, Capaccioli \& D'Onofrio(1993)]{caon93}Caon~N., Capaccioli~M., D'Onofrio~M., 1993, MNRAS, 265, 1013
\bibitem[Cole et al.(2001)]{cole01}Cole~S. et al. (The 2dFGRS Team),
2001, astro-ph/0012429 (submitted to MNRAS)
\bibitem[Cox(2000)]{cox00}Cox~A.~N., 2000, Allen's astrophysical
quantities (4$^{\rm th}$ edition; New York, Springer)
\bibitem[de Jong(1996)]{dejong96}de Jong~R.~S., 1996, A\&AS, 118, 557
\bibitem[de Vaucouleurs(1959)]{devac59}de Vaucouleurs~G., 1959, in Fl\"ugge S., ed., Handbuch der Physik 53, Springer-Verlag, Berlin, p.~275
\bibitem[Dressler(1980)]{dressler80}Dressler~A., ApJ, 236, 351
\bibitem[Efstathiou, Ellis \& Peterson(1988)]{eep}Efstathiou~G., Ellis~R.~S., Peterson~B.~A., 1988, MNRAS, 232, 431 (EEP)
\bibitem[Eggen, Lyden-Bell \& Sandage(1962)]{els}Eggen~O.~J., Lynden-Bell~D., Sandage~A.~R., 1962, ApJ, 136, 748
\bibitem[Fukugita, Hogan \& Peebles(1998)]{fuku98}Fukugita~M., Hogan~C.~J., Peebles~P.~J.~E., 1998, ApJ, 503, 518
\bibitem[Gardner et al.(1997)]{gardner97}Gardner~J.~P., Sharples~R.~M., Frenk~C.~S., Carrasco~B.~E., 1997, ApJ, L99
\bibitem[Hoyle(1949)]{hoyle49}Hoyle~F., 1949, in Burgers~J.~M., van de Hulst~H.~C., eds., Problems of Cosmical Aerodynamics, Central Air Documents Office, Dayton, p.~195
\bibitem[Hubble(1926)]{hubble26}Hubble~E., 1926, ApJ, 64, 321
\bibitem[Jimenez et al.(1999)]{jimenez99}Jimenez~R., Friaca~A.~C.~S., Dunlop~J.~S., Terlevich~R.~J., Peacock~J.~A., Nolan~L.~A., 1999, MNRAS, 305, 16
\bibitem[Kauffmann, White \& Guiderdoni(1993)]{kauff93}Kauffmann~G., White~S.~D.~M., Guiderdoni~B., 1993, MNRAS, 264, 201
\bibitem[Kauffmann(1995, 1996)]{kauff9596}
\vspace{-10mm}\bibitem[Kauffmann(1995)]{kauff95}Kauffmann~G., 1995, MNRAS, 274, 161
\bibitem[Kauffmann(1996)]{kauff96}Kauffmann~G., 1996, MNRAS, 281, 487
\bibitem[Kennicutt(1983)]{k83}Kennicutt~R.~C., 1983, ApJ, 272, 54
\bibitem[Loveday et al.(1992)]{loveday92}Loveday~J., Peterson~B.~A., Efstathiou~G., Maddox~S.~J., 1992, ApJ, 390, 338
\bibitem[Menanteau et al.(1999)]{menanteau99}Menanteau~F., Ellis~R.~S., Abraham~R.~G., Barger~A.~J., Cowie~L.~L., 1999, MNRAS, 309, 208
\bibitem[Naim et al.(1995)]{naim95}Naim~A., Lahav~O., Buta~R.~J., Corwin~H.~G., de Vaucoulers~G., Dressler~A., Huchra~J.~P., van den Bergh~S., Raychaudhury~S., Sodr\'e~L., Storrie-Lombardi~M.~C., 1995, MNRAS, 274, 1107
\bibitem[Navarro, Frenk \& White(1995)]{nfw95}Navarro~J.~F., Frenk~C.~S., White~S.~D.~M., 1995, MNRAS, 275, 56
\bibitem[Navarro \& Steinmetz(1997, 2000)]{navsteindummy}
\vspace{-10mm}\bibitem[Navarro \& Steinmetz(1997)]{navstein97}Navarro~J.~F., Steinmetz~M., 1997, ApJ, 478, 13
\bibitem[Navarro \& Steinmetz(2000)]{navstein00}Navarro~J.~F., Steinmetz~M., 2000, ApJ, 538, 477
\bibitem[Peebles(1969)]{peebles69}Peebles~P.~J.~E., 1969, ApJ, 155, 393
\bibitem[Sandage, Tammann \& Yahil(1979)]{STY}Sandage~A., Tammann~G.~A., Yahil~A., 1979, ApJ, 232, 352
\bibitem[Salpeter(1955)]{sal}Salpeter~E.~E., 1955, ApJ, 121, 61
\bibitem[Schechter(1976)]{schecht76}Schechter~P., 1976, ApJ, 203, 557
\bibitem[Schechter \& Dressler(1987)]{schecht87}Schechter~P.~L., Dressler~A., 1987, AJ, 94, 563
\bibitem[Somerville, Primack \& Faber(2001)]{somerville01}Somerville~R.~S., Primack~J.~R., Faber~S.~M., 2001, MNRAS, 320, 504
\bibitem[Toomre(1977)]{toomre77}Toomre~A., 1977, in Tinsley~B.~M., Larson~R.~B., eds., The Evolution of Galaxies and Stellar Populations. Yale Univ. Press, New Haven, p.~401
\bibitem[van den Bosch, Burkert \& Swaters(2001)]{bosch01}van den Bosch~F.~C., Burkert~A., Swaters~R.~A., 2001, astro-ph/0105158
\bibitem[Wadadekar, Robbason \& Kembhavi(1999)]{wad99}Wadadekar~Y., Robbason~B., Kembhavi~A., 1999, AJ, 117, 1219
\end{thebibliography}
\end{document}